\documentclass[conference]{IEEEtran}
\IEEEoverridecommandlockouts
\usepackage{cite}
\usepackage{amsmath,amssymb,amsfonts}
\usepackage{algorithmic}
\usepackage{graphicx}
\usepackage{textcomp}
\usepackage{xcolor}
\def\BibTeX{{\rm B\kern-.05em{\sc i\kern-.025em b}\kern-.08em
    T\kern-.1667em\lower.7ex\hbox{E}\kern-.125emX}}

\usepackage{subfigure}
\usepackage{enumerate}
\usepackage[ruled]{algorithm2e}
\usepackage{amsthm}
\usepackage{booktabs}
\usepackage{color}
\usepackage{bbding}
\usepackage{multirow,tabularx,bm,enumitem,kantlipsum}

\begin{document}

\title{Multi-Modality is All You Need for Transferable Recommender Systems
}
\author{
\IEEEauthorblockN{
Youhua Li\IEEEauthorrefmark{1}\IEEEauthorrefmark{9} \thanks{\IEEEauthorrefmark{1} This paper was partly done during internship at Westlake University.}, 
Hanwen Du\IEEEauthorrefmark{2}, 
Yongxin Ni\IEEEauthorrefmark{2}, 
Pengpeng Zhao\IEEEauthorrefmark{3}, 
Qi Guo \IEEEauthorrefmark{4}\IEEEauthorrefmark{6}, 
Fajie Yuan \IEEEauthorrefmark{2}\IEEEauthorrefmark{6}\thanks{\IEEEauthorrefmark{6} Equal Correspondence to: Qi Guo and Fajie Yuan.}, 
 Xiaofang Zhou\IEEEauthorrefmark{5}
}
\IEEEauthorblockA{
\IEEEauthorrefmark{1}ShanghaiTech University, Shanghai, China\\ 
\IEEEauthorrefmark{9}Shanghai Innovation Center for Processor Technologies, SHIC \\
\IEEEauthorrefmark{2}Westlake University, Hangzhou, China\\ \IEEEauthorrefmark{3}School of Computer Science and Technology, Soochow University, Suzhou, China\\
\IEEEauthorrefmark{4}State Key Lab of Processors, Institute of Computing Technology, Chinese Academy of Sciences, Beijing, China\\
\IEEEauthorrefmark{5}The Hong Kong University of Science and Technology, Hong Kong SAR, China\\
}
\IEEEauthorblockA{liyh5@shanghaitech.edu.cn, \{duhanwen,niyongxin,yuanfajie\}@westlake.edu.cn\\ ppzhao@suda.edu.cn, guoqi@ict.ac.cn, zxf@cse.ust.hk 
}
}

\maketitle
\IEEEpeerreviewmaketitle
\begin{abstract}

ID-based Recommender Systems (RecSys), where each item is assigned a unique identifier and subsequently converted into an embedding vector, have dominated the designing of RecSys. Though prevalent, such ID-based paradigm is not suitable for developing transferable RecSys and is also susceptible to the cold-start issue. In this paper, we unleash the boundaries of the ID-based paradigm and propose a Pure Multi-Modality based Recommender system (PMMRec), which relies solely on the multi-modal contents of the items (e.g., texts and images) and learns transition patterns general enough to transfer across domains and platforms. Specifically, we design a plug-and-play framework architecture consisting of multi-modal item encoders, a fusion module, and a user encoder. To align the cross-modal item representations, we propose a novel next-item enhanced cross-modal contrastive learning objective, which is equipped with both inter- and intra-modality negative samples and explicitly incorporates the transition patterns of user behaviors into the item encoders. To ensure the robustness of user representations, we propose a novel noised item detection objective and a robustness-aware contrastive learning objective, which work together to denoise user sequences in a self-supervised manner. PMMRec is designed to be loosely coupled, so after being pre-trained on the source data, each component can be transferred alone, or in conjunction with other components, allowing PMMRec to achieve versatility under both multi-modality and single-modality transfer learning settings. Extensive experiments on 4 sources and 10 target datasets demonstrate that PMMRec surpasses the state-of-the-art recommenders in both recommendation performance and transferability. Our code and dataset is available at: https://github.com/ICDE24/PMMRec.

\end{abstract}
 \section{Introduction}
 \begin{figure}[htbp]
	\begin{center}
    \includegraphics[width=1.0\linewidth]{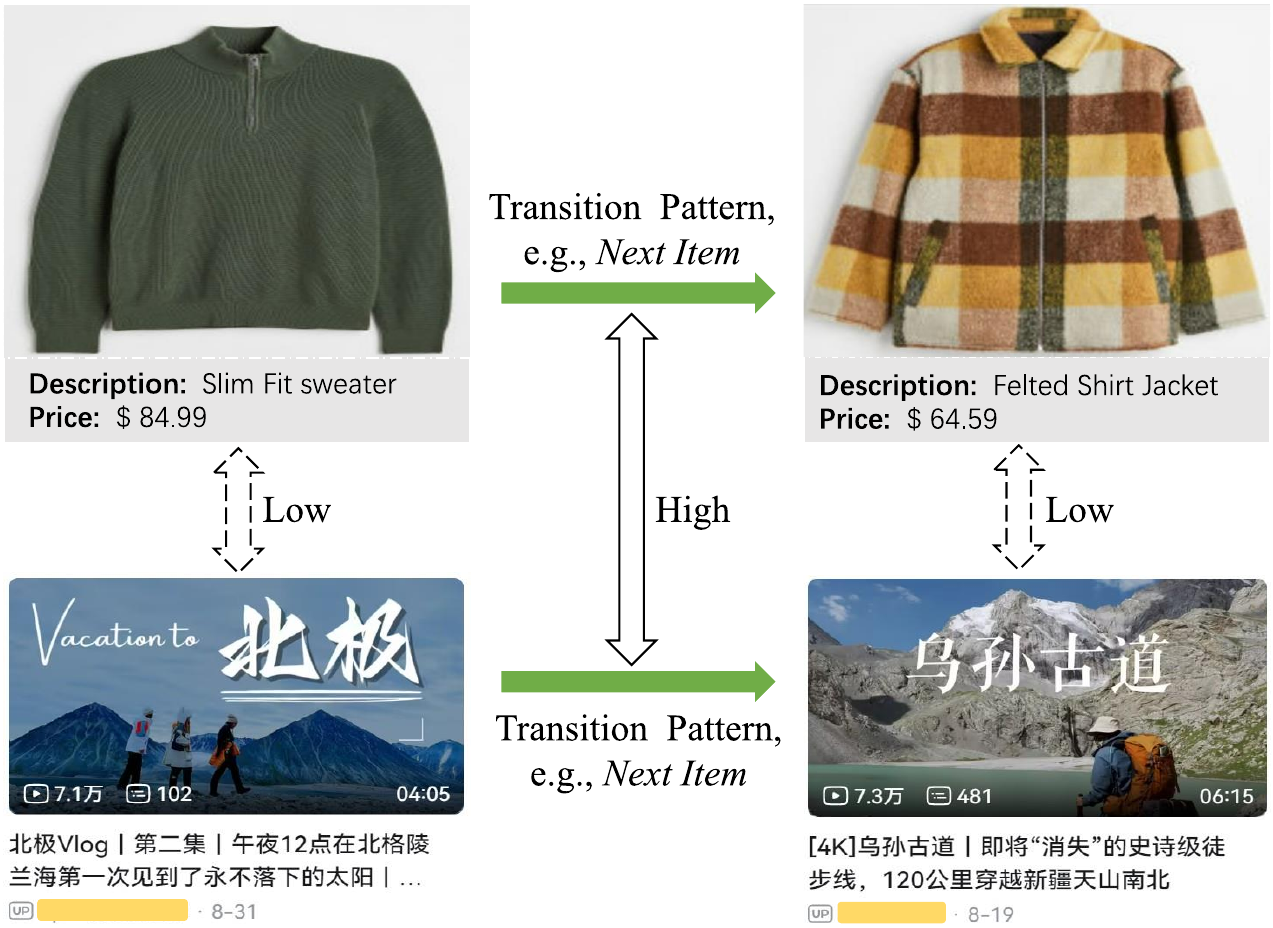}
    \end{center}
    \caption{An example from the HM dataset and the Bili dataset. Although the content similarities between different platforms might be low, the commonalities of universal transition patterns (e.g., next-item transition) between different platforms are still high, making it beneficial to transfer knowledge across different domains and platforms. Top: the HM dataset (online e-commerce). Bottom: the Bili dataset (short-video sharing).}
    \label{sequential}
\end{figure}

Recommender Systems (RecSys) have become one of the most useful information filtering engines for online applications (e.g., e-commerce and  short-video sharing). In recent years, sequential recommendation, which aims to capture the user's dynamic preference and recommend a user with an unseen item that this user is likely to interact in the future, becomes an important task of RecSys with growing research interest. Typical sequential recommendation methods assign a unique identifier (ID) for each item in the dataset, which is subsequently converted into an embedding vector such that a sequence encoder can extract user preferences from the sequence of item embeddings. Framed under this \textit{ID-based} paradigm, various types of sequence encoders, such as Recurrent Neural Networks (RNNs) \cite{hidasi2015session}, Convolutional Neural Networks (CNNs) \cite{yuan2019simple}, and Transformers \cite{kang2018self,sun2019bert4rec}, have been successfully applied in sequential recommendation.

Despite remarkable successes and widespread applications, such ID-based paradigm actually has inherent flaws that might hinder the future development of RecSys. The most critical issue (arguably) of this ID-based paradigm is its poor transferability~\cite{wang2022transrec,yuan2023go,cheng2023image,fu2023exploring,li2023exploring,zhang2023ninerec,wang2023missrec,yang2023collaborative,ni2023contentdriven}. Users' interaction data on different domains and platforms, though dissimilar considering their content differences, should share some common knowledge and transition patterns (e.g., next-item transition) transferable across domains and platforms, as is illustrated in Figure \ref{sequential}. However, as item IDs are not shareable across platforms, transferring common knowledge from one platform to the other becomes almost impossible for ID-based recommenders. Moreover, since the ID embeddings of the items cannot be well-trained with only limited interaction data \cite{zhu2021learning,zhao2022improving}, ID-based recommenders are prone to suffer from the cold-start issue \cite{dong2020mamo,hou2022towards,wei2021contrastive} when fewer interaction data are available.

\textbf{Is it possible to unleash the boundaries of the ID-based paradigm, and usher in a new paradigm for RecSys that is more transferable across platforms and less susceptible to the cold-start issue?} In this paper, we shift our focus from the conventional ID-based paradigm and resort to the multi-modal contents of the item itself. In fact, our experience of the world is multi-modal (i.e., perceptible from multiple media such as texts and images), and in order for RecSys to better understand our preferences, it also needs to interpret users' interactions from a multi-modal perspective.  
Therefore, we optimistically envision a whole new paradigm for the next generation of recommender system that, instead of taking item IDs as input, relies solely on the multi-modality contents of the items. Building such a pure multi-modality based RecSys has very desirable properties: first, RecSys trained with pure multi-modality data are naturally endowed with strong transferability, since different items on different platforms will be represented as unified forms of texts and/or images, and therefore can be jointly processed under a unified framework without the need for item IDs; second, multi-modal contents shed light on the diverse transition patterns (e.g., complex transitions between image style, brand category and text description) otherwise unobserved in the ID-based interaction data, which can be very helpful for alleviating the cold-start issue \cite{zhang2019feature}; third, ongoing researches under the pure multi-modality based recommendation paradigm can directly benefit from technical breakthroughs in Natural Language Processing (NLP) and Computer Vision (CV) (e.g., utilize the state-of-the-art pre-trained language or vision models to extract more informative multi-modal item representations), serving as the bond of the NLP, CV and RecSys community, and further paving the way for the development of generic foundation models \cite{singh2022flava,wang2022transrec,li2023exploring,cheng2023image} for RecSys in the future.

Though with promising applications, developing a pure multi-modality based RecSys actually faces serious obstacles, and several open challenges remain to be solved:

\textbf{Representation Alignment.} 
In our pure multi-modality setting, images and texts are first encoded independently by the item encoders with different architectures and then processed by the user encoder for the recommendation task. To learn transferable representations for the recommendation task, two types of representation alignment must be achieve: \textit{(a) Representation Alignment between Multi-modalities.} Given that texts and images are independently encoded by different encoders (e.g., RoBERTa \cite{liu2019roberta} and Vision Transformer \cite{dosovitskiy2020image}), the representation space of texts and images might be different. To jointly process texts and images in the same representation space, representation alignment between multi-modalities must be achieved. \textit{(b) Representation Alignment in Recommendation Semantic Space.} Even if the representation space of texts and images are well-aligned, these representations only characterize the semantics of texts or the patterns of images that are not suitable for the recommendation task. To endow the multi-modality representations with meaningful information specific to the recommendation task, representation alignment in recommendation semantic space must be achieved. In the literature, transferable recommendation methods such as  \cite{hou2022towards,hou2022learning}  adopt parametric whitening or vector quantization to obtain isotropic text representations, yet such alignment techniques are designed for the text modality only. Therefore, how to design proper techniques that effectively align multi-modal representations remains to be solved.

\textbf{Severe Data Noises.} The development of RecSys has long been troubled by the data noise issue, since ratings, reviews or interactions recorded by online platforms may not truly reflect the user's intentions. In our multi-modality setting, the data noise issue is exacerbated, where the contents of texts and images tend to contain noises or mismatches with the recommendation setting (e.g., the text of a item is inconsistent with the image of this item). In the literature, some methods \cite{qin2021world,gao2022self} learn to denoise implicit feedbacks for recommendation via a two-stage optimization strategy, but these approaches are still limited under the ID-based paradigm and thus not suitable for our pure multi-modality setting. Therefore, how to design proper techniques to remedy the adverse effects of the noisy data is still an open question.

\textbf{Versatility.} In real-world applications, transferable RecSys should be versatile enough such that it can be flexibly deployed in various scenarios. For example, after being pre-trained on a multi-modal dataset with both text and image, sometimes we may wish to deploy our model on downstream platforms where only a single modality (text or image) is available. To enable such versatility, the architecture of the pure multi-modality RecSys must be flexible enough to adapt for various transfer learning settings (e.g., text+image $\rightarrow$ text only, text+image $\rightarrow$ image only), and how to design such flexible architecture remains an open challenge. 

To tackle the challenged mentioned above, we propose a \textbf{P}ure \textbf{M}ulti-\textbf{M}odality based \textbf{Rec}ommender system (PMMRec), a pure multi-modality based transferable sequential recommender capable of achieving recommendation performances superior or on par with ID-based sequential recommenders. Our framework designs multiple plug-and-play components with effective representation alignment and denoising objectives to enable versatile transferability across domains and platforms. Specifically, we leverage pre-trained text and vision encoders as item encoders to extract the modality-specific feature embeddings. The modality-specific feature embeddings are fused into multi-modal item representation via a fusion module, which is subsequently processed by a user encoder to capture the recommendation transition patterns. 
To achieve representation alignment, we propose a novel next-item enhanced cross-modal contrastive learning objective, which is enhanced with both inter- and intra-modality negatives samples to achieve representation alignment between multi-modalities, and explicitly incorporates transition patterns of user behaviors into the item encoders to achieve representation alignment in the recommendation semantic space. To handle the data noise issue, we propose noised item detection to make the model aware of synthetic noises in the corrupted user sequences, and further enhance framework robustness by contrasting the original user sequences with the corrupted user sequences. After being pre-trained on the source data, the plug-and-play architectures of the item encoders, the multi-modal fusion module and the user encoder can be transferred alone or in conjunction with other components, thus allowing PMMRec to achieve versatility under both multi-modality and single-modality transfer learning settings. 

We have observed that most existing works on transferable recommendation \cite{wang2023missrec,yang2023collaborative,zhang2023multimodal} have primarily been validated in the same or similar scenarios (e.g., Amazon and other online e-commerce platforms with clean or pure backgrounds) \cite{hou2022towards}. To further evaluate the applicability of our framework in real-world settings, we use the data from two publicly available e-commerce datasets (Amazon and HM) and gather additional data from two online short video platforms (Kwai and Bilibili), resulting in \textbf{4} source and \textbf{10} target datasets for conducting comprehensive cross-domain and cross-platform transfer learning experiments. We conduct extensive experiments on these datasets, and our experiment results demonstrate the broad applicability of PMMRec in various real-world recommendation scenarios. 

In summary, our contributions can be outlined as follows:

\begin{itemize}[leftmargin = 8pt, topsep = 1pt]

\item We propose a novel Pure Multi-Modality based Recommender System (PMMRec) that surpasses state-of-the-art recommenders in terms of both recommendation performance and transferability.

\item We design a plug-and-play architecture consisting of item encoders, a multi-modal fusion modal and a user encoder, as the framework backbone, which achieves good synergy for recommendation pre-training and allows for good versatility under various transfer learning settings.

\item We propose a next-item enhanced cross-modal contrastive learning objective to achieve representation alignment, and propose two self-supervised denoising objectives to enhance framework robustness against data noises.

\item We evaluate the performance of PMMRec using the datasets from different domains and platforms under versatile transfer settings. Extensive experiments on real-world datasets demonstrate the broad applicability of PMMRec in various recommendation scenarios. 

\end{itemize}

\section{Related Works}

\subsection{Sequential Recommendation}

Sequential recommendation views the user's interaction history as a chronologically ordered sequence and aims the capture the dynamic transitions of the user's preference.
Traditional methods for sequential recommendation employ Markov Chains (MCs) to model the sequential dynamics of the user's interactions, such as MDP \cite{shani2005mdp}, FPMC \cite{rendle2010factorizing} and Fossil \cite{he2016fusing}. 
With the advancements in deep learning, an ID-based recommendation paradigm, where each item is assigned a unique ID, converted into an embedding vector, and processed by a sequence encoder, has been established. Under this ID-based paradigm, researchers have developed various types of deep neural networks as effective sequence encoders. For example, GRU4Rec \cite{hidasi2015session} introduces RNN with ranking loss functions for session-based recommendation, Caser \cite{tang2018personalized} adopts CNN with horizontal and vertical convolutional layers to capture the diverse behavior patterns for sequential recommendation, NextItNet \cite{yuan2019simple} improves Caser \cite{tang2018personalized} by introducing dilated convolutional layers and residual connections. Recently, the success of Transformers \cite{vaswani2017attention} has motivated the designing of attention-based sequential recommendation methods. For example, SASRec \cite{kang2018self} adopts unidirectional Transformer with the next-item prediction task for sequential recommendation, BERTRec \cite{sun2019bert4rec} employs bidirectional Transformer with the masked item prediction task for sequential recommendation. 

Some works also try to fuse the multi-modal contents of items under the ID-based backbone to improve recommendation performance. For example, FDSA \cite{zhang2019feature} fuses the textual features with ID-based embedding vectors via feature-level self-attention blocks, 
CARCA \cite{rashed2022carca} adopts a cross-attention network to fuse item attributes or image-based features. Despite effectiveness, all these methods rely on the ID-based paradigm to make recommendations, making them difficult to transfer across domains and platforms. 

\subsection{Transferable Recommendation}
Transfer learning has also been proven useful in the development of recommender systems, since it can alleviate the data sparsity issue by transferring knowledge from source domains and platforms to improve recommendation performance on the target domains and platforms. For example, $\pi$-net \cite{ma2019pi} leverages shared user accounts to transfer user information to the target domain, PeterRec \cite{yuan2020parameter} and
Conure~\cite{yuan2021one} learn transferable user representations by pre-training on user's behavior sequences, RecGURU \cite{li2022recguru} employs adversarial learning to learn generalized user representations for cross-domain recommendation. To achieve transfer learning in a more general form, some methods attempt to utilize side features of items as identifiers, which can generalize to different domains and platforms. For example, ZESRec \cite{ding2021zero} employs pre-extracted text embeddings as transferable item representations for cross-task knowledge transfer, 
TransRec~\cite{wang2022transrec,fu2023exploring} and UniSRec \cite{hou2022towards} utilize the associated description text or images of items to learn transferable item representations, VQRec \cite{hou2022learning} improves UniSRec \cite{hou2022towards} by mapping item text to discrete item codes and deriving item representations via the code embedding table. MoRec~\cite{yuan2023go,li2023exploring} empirically demonstrates that, pure modality-based recommender models can perform on par with ID-based recommenders by replacing the ID embeddings with text or vision representations.

Despite effectiveness, these methods cannot utilize both texts and images as multi-modal contents and lack alignment techniques in the representation space, thus they show weaker performance and do not apply to versatile transfer learning settings involving both texts and images.

\subsection{Multi-Modal Learning}
Multi-modal learning leverages data from multiple modalities to understand their intrinsic connections. Various researches have proven the usefulness of learning from multiple modalities~\cite{yang2023collaborative,wang2023missrec}, such as achieving entity alignment via multi-modal knowledge graph \cite{chen2022multimodalsiamesenetwork}, training on massive corpus of multi-modal data for image-text retrieval \cite{radford2021learning}, and building a foundation model for vision and language alignment \cite{singh2022flava}.

In recommender systems, researchers have also tried to incorporate multi-modal data into the recommendation model to provide richer recommendation contexts and improve recommendation performance. For example, VBPR \cite{he2016vbpr} empowers factorization model with visual features to achieve more personalized ranking, MMGCN \cite{MMGCN} proposes a multi-modal graph convolutional network for personalized video recommendation \cite{MMGCN}, LATTICE \cite{zhang2021mining} combines the modality-aware graphs with the initial user-item graph to mine richer semantics for collaborative filtering, DualGNN \cite{wang2021dualgnn} proposes a graph-based multi-modal representation learning framework for micro-video recommendation. 
Different from these works that treat multi-modal features as contextual information, we aim to build a pure multi-modality-based RecSys that relies solely on the multi-modality contents of the items, thus enabling transferability under versatile transfer learning settings and further paving the way for the development foundation models for RecSys in the future.

\section{Framework Architecture}
\begin{figure*}[htbp]
    \begin{center}
        \includegraphics[width=0.96\linewidth]{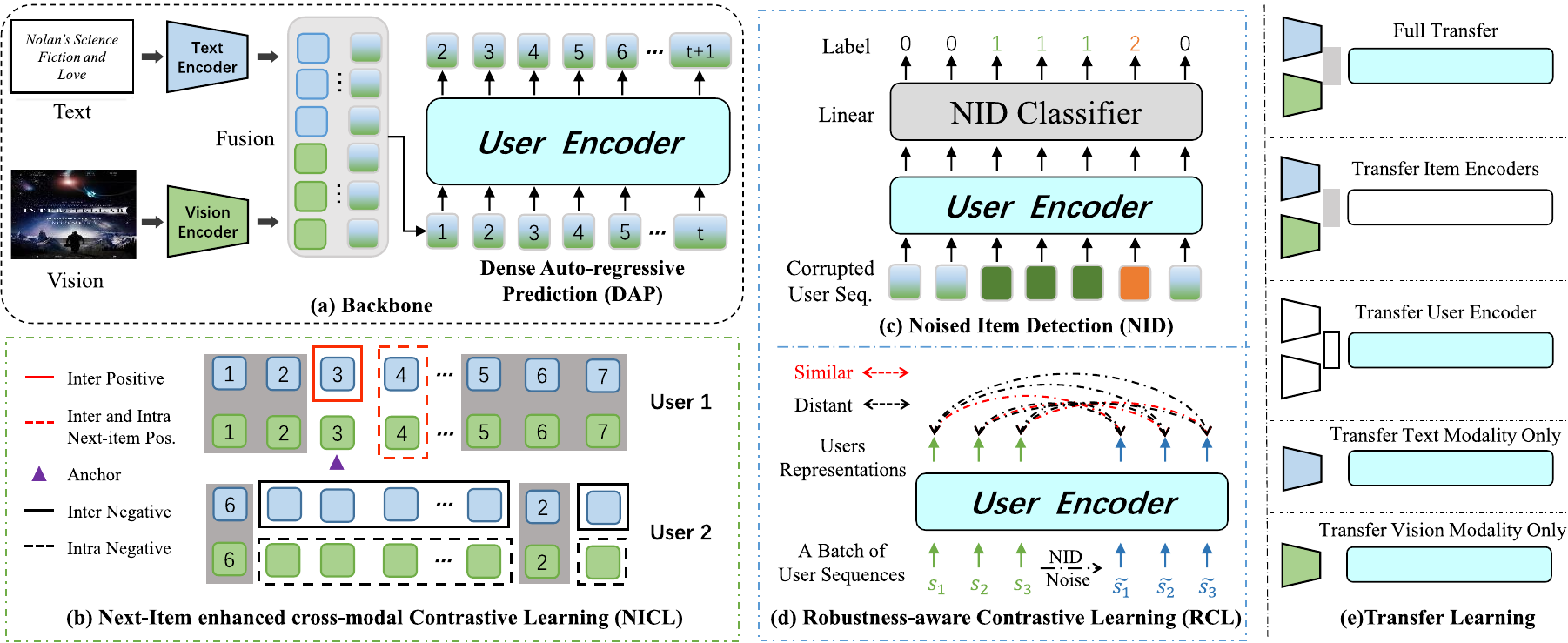}
    \end{center}
        \caption{The architecture of PMMRec. \textbf{(a)}: The backbone of PMMRec framework, including the item encoders (text encoder and vision encoder), the multi-modal fusion module and the user encoder. \textbf{(b)}: NICL achieves representation alignment via next-item enhanced intra- and inter-modality contrastive learning. \textbf{(c)}: NID adapts the user encoder to noisy data by making it aware of synthetic noises. \textbf{(d)}: RCL enhances framework robustness by contrasting the original user sequences with the corrupted user sequences.  \textbf{(e)}: PMMRec achieves versatility under various transfer learning settings.}
    \label{overview}
    \vspace{-0.1cm}
\end{figure*}

In this section, we present the framework architecture of PMMRec. We start by introducing the problem statement of transferable recommendation in Section~\ref{problemstatement}. Next, we introduce the backbone of our framework in Section~\ref{Backbone}. Moreover, we introduce the objective functions for framework pre-training, including the next-item enhanced cross-modal contrastive learning objective in Section~\ref{NICL}, and the self-supervised denoising objective in Section~\ref{selfsupervisedenoising}. Finally, we illustrate how our framework achieves versatile transfer  in Section~\ref{tranferlearning}. Figure~\ref{overview} presents an overview of our framework.
\subsection{Problem Statement}\label{problemstatement}
Let $\mathcal{U}$ denote a set of users and $\mathcal{I}$ denote a set of items. For each user $u\in\mathcal{U}$, we define $s_u=(i^{u}_1,i^{u}_2,\cdots,i^{u}_L)$ as this user's interaction sequence sorted in the chronological order, where each item $i^{u}_l{\in}\thinspace\mathcal{I}\thinspace(1{\leq}l{\leq}L)$ in this sequence is represented not by its item ID, but by its text and vision features $\{T_{i^{u}_l},V_{i^{u}_l}\}$. For transferable recommendation, we consider the classical setting where the model is first pre-trained on the source dataset and then fine-tuned on the target dataset. Different from traditional cross-domain recommendations, our method does not require any overlapping users or items between the source and target domain or platform.
\subsection{Backbone}
\label{Backbone}
In this subsection, we introduce several essential modules in the PMMRec backbone, as is illustrated in Figure \ref{overview}(a). For an item with text and vision contents, we employ the corresponding pre-trained encoders to learn its text and vision features. Next, we propose a multi-modal fusion module to combine modality-specific feature embeddings into a cross-modal representation of the item. Finally, the cross-modal item representation is fed into a user encoder to capture the transition patterns for the recommendation task. 
\subsubsection{Text Encoder}
The Text Encoder (TE) is responsible for generating high-quality text embeddings that effectively capture the semantics of an item's text. In this paper, we adopt the multilingual RoBERTa\cite{liu2019roberta}, a state-of-the-art pre-trained language model, as our default TE. Specifically, given an item $i^{u}_{l}$ and its corresponding textual contents $T_{i^{u}_{l}}=[token_{i^{u}_{l}}^{1},token_{i^{u}_{l}}^{2},\cdots,token_{i^{u}_{l}}^{p}]$ with length $p$, we concatenate (1) a special symbol $token^{cls}_{i^{u}_{l}}$ (2) the tokens of the item text and feed the concatenated sequence into the TE model, which outputs a final hidden vector for each input token. The encoding process of the item texts is formulated as follows:
\begin{equation}
(\mathbf{t}_{i^{u}_{l}}^{cls}; \mathbf{t}_{i^{u}_{l}}^{1}, \dots, \mathbf{t}_{i^{u}_{l}}^{p}) = TE([token^{cls}_{i^{u}_{l}}; token_{i^{u}_{l}}^{1}, \dots, token_{i^{u}_{l}}
^{p}])
\end{equation}
where $\mathbf{t}_{i^{u}_{l}}^{cls}, \mathbf{t}_{i^{u}_{l}}^{1},\dots,\mathbf{t}_{i^{u}_{l}}^{p}\in \mathbb{R}^d$ represent the output hidden vectors corresponding to each input token, $d$ is the hidden dimensionality of text encoder. The $\mathbf{t}^{cls}_{i^{u}_{l}}$ represents the text-modality feature embedding of this item, which will be used for cross-modal representation alignment objectives, and all the $\mathbf{t}_{i^{u}_{l}}^{1}, \dots, \mathbf{t}_{i^{u}_{l}}^{p}$ are used for multi-modal fusion.

\subsubsection{Vision Encoder} 
We apply Vision Transformers (ViT) \cite{dosovitskiy2020image} as Vision Encoder (VE) to ensure homogeneity with the TE. We select the representative CLIP-ViT \cite{radford2021learning} as our default VE. It divides input images into multiple fixed-size patches and projects each patch into a fixed-length vector, then a special symbol $patch^{cls}_{i^{u}_{l}}$ is added to the input patches sequence, both of which are fed into the ViT. Given an item $i^{u}_{l}$ and its corresponding image features (segmented into image patches) $V_{i^{u}_{l}}=[patch_{i^{u}_{l}}^{1},patch_{i^{u}_{l}}^{2},\cdots,patch_{i^{u}_{l}}^{q}]$ with length $q$, the encoding process of item image is formulated as follows:
\begin{equation}(\mathbf{v}_{i^{u}_{l}}^{cls}; \mathbf{v}_{i^{u}_{l}}^1, \dots, \mathbf{v}_{i^{u}_{l}}^{q}) = VE([patch^{cls}_{i^{u}_{l}}; patch_{i^{u}_{l}}^{1}, \dots, patch_{i^{u}_{l}}^{q}])
\end{equation}
where $\mathbf{v}_{i^{u}_{l}}^{cls},\mathbf{v}_{i^{u}_{l}}^{1},\dots,\mathbf{v}_{i^{u}_{l}}^{q}\in \mathbb{R}^d$ represent the output hidden vectors corresponding to each input patch, $d$ is the hidden dimensionality kept the same as the TE. The $\mathbf{v}_{i^{u}_{l}}^{cls}$ represents the vision-modality feature embedding of this item, which will be used for cross-modal representation alignment objectives, and all the $\mathbf{v}_{i^{u}_{l}}^1, \dots, \mathbf{v}_{i^{u}_{l}}^{q}$ are used for multi-modal fusion.

\subsubsection{Multi-modal Fusion}
As a key module in multi-modal architectures, we employ a Transformer-based merge-attention method that is powerful and efficient for multiple modalities. Specifically, it concatenates text tokens and visual patches together and feeds them into a single fusion block. Then a multi-modal special symbol $\mathbf{mm}^{cls}_{i^{u}_{l}}\in\mathbb{R}^d$ is added to the concatenation of text tokens and image patches to form the multi-modal representation sequence of this item. Finally, the multi-modal representation sequence is fed into a Transformer layer and the corresponding output of the $\mathbf{mm}^{cls}_{i^{u}_{l}}$ is fetched as the final representation for global multi-modal semantics:
\begin{equation}
(\mathbf{e}^{cls}_{i^{u}_{l}};\dots;\dots)=Fuse([\mathbf{mm}^{cls}_{i^{u}_{l}}; \mathbf{t}_{i^{u}_{l}}^{1}, \dots, \mathbf{t}_{i^{u}_{l}}^{p}; \mathbf{v}_{i^{u}_{l}}^{1}, \dots, \mathbf{v}_{i^{u}_{l}}^{q}])
\end{equation}
where $\mathbf{e}^{cls}_{i^{u}_{l}} \in \mathbb{R}^d$ represents the final multi-modal representation of this item. 

 \subsubsection{User Encoder} To capture the transition patterns for the recommendation task, we need a recommendation network to model the user's interaction sequence based on learned cross-modal item representations, which we refer to as the \textit{User Encoder}. In this paper, we directly adopt Transformer encoder as the user encoder (denoted as $\mathrm{Trm}$), which is kept the same as SASRec \cite{kang2018self} for a fair comparison. The user encoder takes in a sequence of the cross-modal representations of items as input, and outputs its hidden representations:
\begin{equation}\label{userencoder}
(\mathbf{h}_{i^{u}_{1}},\mathbf{h}_{i^{u}_{2}},\dots,\mathbf{h}_{i^{u}_{L}})=\mathrm{Trm}(\mathbf{e}^{cls}_{i^{u}_{1}}+\mathbf{p}_{1},\mathbf{e}^{cls}_{i^{u}_{2}}+\mathbf{p}_{2},\dots,\mathbf{e}^{cls}_{i^{u}_{L}}+\mathbf{p}_{L})
\end{equation}

where $\mathbf{h}_{i^{u}_{l}}\in\mathbb{R}^{d}\thinspace(1{\leq}l{\leq}L)$ denotes the hidden representation of the $l$-th item in the user's interaction sequence, $\mathbf{p}_{l}\in\mathbb{R}^{d}\thinspace(1{\leq}l{\leq}L)$ denotes the learnable position embedding vector for the $l$-th position. For a batch of user sequences $\{s_u\}^{B}_{u=1}$ with batch size $B$, the user encoder is optimized via Dense Auto-regressive Prediction (DAP), which requires the model to predict the next item based on previous interactions:
\begin{equation}\label{DAP}
\begin{aligned}
\mathcal{L}^{DAP}_{i^{u}_{l}}=-&\log \frac{\exp(\mathbf{h}^{\top}_{i^{u}_{l}}{\cdot}\mathbf{e}^{cls}_{i^{u}_{l+1}})}{\exp(\mathbf{h}^{\top}_{i^{u}_{l}}{\cdot}\mathbf{e}^{cls}_{i^{u}_{l+1}}) +\sum\limits_{i^{k}_{j}\in\mathcal{N}_{i^{u}_{l}}} \exp(\mathbf{h}^{\top}_{i^{u}_{l}}{\cdot}\mathbf{e}^{cls}_{i^{k}_{j}}) }\\
&\mathcal{L}^{DAP}=\sum\limits_{u=1}^{B}\sum\limits_{l=1}^{L-1}\frac{1}{B*(L-1)}\mathcal{L}^{DAP}_{i^{u}_{l}}
\end{aligned}
\end{equation}
where $\mathcal{N}_{i^{u}_{l}}\!=\!\{i^{k}_{j}\thinspace|\thinspace{\forall}{\thinspace}1{\leq}k{\leq}B,k{\neq}u\}$ denotes the set of in-batch negative samples for item $i^{u}_{l}$, which contains all the items in the batch except the interacted items of the current user. So long as the training data come from more than one source (domain or platform), negatives will be sampled from different sources, which helps the model to recognize different item styles and develop generalized user representations.

\subsection{Next-item enhanced cross-modal Contrastive Learning}\label{NICL}
To achieve cross-modal representation alignment, we start by introducing the Vanilla cross-modal Contrastive Learning (VCL) objective. Next, we propose an intra-modality negative sampling strategy to improve VCL, resulting in the Intra-modality sample enhanced cross-modal Contrastive Learning (ICL) objective. Finally, based on VCL and ICL, we propose a novel Next-Item enhanced cross-modal Contrastive Learning (NICL), which is enhanced with next-item positive samples to achieve representation alignment in the recommendation semantic space. 
\subsubsection{Vanilla Cross-modal Contrastive Learning}
VCL  utilizes self-supervised learning objectives to align item texts and images within a shared space.
By maximizing the similarity between positive pairs and minimizing the similarity between negative pairs, the model learns similar representations for matching modalities and distinct ones for non-matching modalities. The VCL objective is defined as follows:
\begin{equation}
\begin{aligned}
\mathcal{L}_{i^{u}_{l}}^{T,V} &= -\log \frac{\delta(t^{u}_{l}, v^{u}_{l})}{\delta(t^{u}_{l}, v^{u}_{l}) + \sum\limits_{i^{k}_{j}\in\mathcal{N}_{i^{u}_{l}}} \delta(t^{u}_{l}, v^{k}_{j})}\\  
\mathcal{L}_{i^{u}_{l}}^{V,T} &= -\log \frac{\delta(v^{u}_{l}, t^{u}_{l})}{\delta(v^{u}_{l}, t^{u}_{l}) + \sum\limits_{i^{k}_{j}\in\mathcal{N}_{i^{u}_{l}}} \delta(v^{u}_{l}, t^{k}_{j}) }
\end{aligned}
\end{equation}
where $\delta(t^{u}_l, v^{u}_l) = \exp({\mathbf{t}_{i^{u}_{l}}^{cls}}^{\top}{\cdot}\mathbf{v}_{i^{u}_{l}}^{cls})$ (and similarly with $\delta(v^{u}_l, t^{u}_l)$) denotes the exponential of the dot-product similarity between a positive pair of text-modality and vision-modality feature embeddings, $\mathcal{N}_{i^{u}_{l}}$ (same as the negative samples of another modality in Eq.~\ref{DAP}) denotes the set of in-batch negative samples for item $i^{u}_{l}$.
Note that before computing the similarity, we apply $\ell_2$-normalization to the modality-specific feature embeddings for regularization. 
\subsubsection{Intra-modality Sample Enhancement}
VCL mainly aligns different modalities, but undervalues intra-modality alignment. 
Therefore, we emphasize the importance of intra-modal alignment and introduce supplementary negative samples to address this issue. Compared to standard approaches that only consider inter-modality alignment, our approach diversifies negative samples and augments the overall alignment efficacy. 
The ICL objective is defined as follows:
\begin{equation}
\begin{aligned}
\mathcal{L}_{i^{u}_{l}}^{T,V} &= -\log \frac{\delta(t^{u}_{l}, v^{u}_{l})}{\delta(t^{u}_{l}, v^{u}_{l}) + \sum\limits_{i^{k}_{j}\in\mathcal{N}_{i^{u}_{l}}} \delta(t^{u}_{l}, v^{k}_{j})+ \sum\limits_{i^{k}_{j}\in\mathcal{N}_{i^{u}_{l}}} \delta(t^{u}_{l}, t^{k}_j)}\\
\mathcal{L}_{i^{u}_{l}}^{V,T} &= -\log \frac{\delta(v^{u}_{l}, t^{u}_{l})}{\delta(v^{u}_{l}, t^{u}_{l}) + \sum\limits_{i^{k}_{j}\in\mathcal{N}_{i^{u}_{l}}}\delta(v^{u}_{l}, t^{k}_{j})+ \sum\limits_{i^{k}_{j}\in\mathcal{N}_{i^{u}_{l}}}\delta(v^{u}_{l}, v^{k}_j)}
\end{aligned}
\end{equation}
where $\delta(t^{u}_l, t^{k}_j)=\exp( {\mathbf{t}_{i^{u}_{l}}^{cls}}^{\top}{\cdot}\mathbf{t}_{i^{k}_{j}}^{cls})$ (and similarly with $\delta(v^{u}_l, v^{k}_j)$) denotes the pair of intra-modality negative samples.

\subsubsection{Positive Enhancement by Inter- and Intra-modality Next Items} Although the representations from different modalities can be effectively aligned via ICL, they lack meaningful semantics specific to the recommendation task. To achieve representation alignment in the recommendation semantic space, we propose a novel Next-Item enhanced cross-modal Contrastive Learning (NICL) objective by explicitly incorporating transition patterns of user behaviors into the item encoders. Specifically, NICL further enhances ICL with both inter- and intra-modality feature embeddings of the next item as positive samples, thus granting the item encoders the ability to predict the next item even without accessing user-item interactions. 
As is illustrated in Figure \ref{overview}(b), for the anchor "image-3", we take not only "text-3" but also  "image-4" and "text-4" as the positive samples, and regard all items except "3" and "4" from other users as negative samples. The NICL objective for each modality is formulated as follows:
\begin{equation}
\label{NICL_two_modalities}
\begin{aligned}
\mathcal{L}^{T,V}_{i^{u}_{l}}\!&= -\log \frac{\delta(t^{u}_{l}, v^{u}_{l}) + \delta(t^{u}_{l}, v^{u}_{l+1}) + \delta(t^{u}_l, t^{u}_{l+1})}{\delta(t^{u}_l, v^{u}_l) + \sum\limits_{i^{k}_{j}\in\mathcal{N}_{i^{u}_{l}}}\delta(t^{u}_{l}, v^{k}_j) + \sum\limits_{i^{k}_{j}\in\mathcal{N}_{i^{u}_{l}}}\delta(t^{u}_{l}, t^{k}_j)}\\
\mathcal{L}_{i^{u}_{l}}^{V,T}\!&=-\log\frac{\delta(v^{u}_{l}, t^{u}_{l}) + \delta(v^{u}_{l}, t^{u}_{l+1}) + \delta(v^{u}_l, v^{u}_{l+1})}{\delta(v^{u}_l, t^{u}_l) + \sum\limits_{i^{k}_{j}\in\mathcal{N}_{i^{u}_{l}}}\delta(v^{u}_{l}, t^{k}_j) + \sum\limits_{i^{k}_{j}\in\mathcal{N}_{i^{u}_{l}}}\delta(v^{u}_{l}, v^{k}_j)}
\end{aligned}
\end{equation}

where $\delta(t^{u}_l, v^{u}_{l+1})=\exp( {\mathbf{t}_{i^{u}_{l}}^{cls}}^{\top}{\cdot}\mathbf{v}_{i^{u}_{l+1}}^{cls})$ (and similarly with $\delta(v^{u}_l, t^{u}_{l+1})$) denotes the intra-modality positive pair of item $i^{u}_{l}$ and the next item $i^{u}_{l+1}$,  $\delta(t^{u}_l, t^{u}_{l+1})=\exp( {\mathbf{t}_{i^{u}_{l}}^{cls}}^{\top}{\cdot}\mathbf{t}_{i^{u}_{l+1}}^{cls})$ (and similarly with $\delta(v^{u}_l, v^{u}_{l+1})$) denotes the inter-modality positive pair of item $i^{u}_{l}$ and the next item $i^{u}_{l+1}$. Based on Eq.~\ref{NICL_two_modalities}, we can obtain the total NICL objective, which is calculated twice for both text and vision modalities to achieve symmetry:
\begin{equation}
    \mathcal{L}^{NICL} =  \frac{1}{B*(L-1)}\sum\limits_{u=1}^{B}\sum\limits_{l=1}^{L-1}(\mathcal{L}^{T,V}_{i^{u}_{l}}+\mathcal{L}_{i^{u}_{l}}^{V,T})/2 
\end{equation}

\subsection{Self-supervised Denoising}\label{selfsupervisedenoising}
To make our framework more robust to data noises, we propose two self supervised denoising objectives: (1) a noised item detection objective to adapt the user encoder to data noises; (2) a robustness-aware contrastive learning objective to further enhance framework robustness against data noises.
\subsubsection{Noised Item Detection} 
Drawing inspiration from \cite{clark2020electra}, we design Noised Item Detection (NID) to adapt the user encoder to noisy data, as is shown in Figure~\ref{overview}(c). For each user sequence $s_u$, we obtain the corrupted user sequence $\tilde{s}_{u}$ by (1) shuffling $15\%$ of the items in the sequence (2) replacing an additional $5\%$ of the items in the sequence with random items from the batch. We then feed the corrupted user sequence $\tilde{s}_{u}$ into the user encoder (Eq.~\ref{userencoder}) and generate the corresponding hidden representations $(\tilde{\mathbf{h}}_{i^{u}_{1}},\tilde{\mathbf{h}}_{i^{u}_{2}},\dots,\tilde{\mathbf{h}}_{i^{u}_{L}})$, which is subsequently converted into logits via the linear NID classifier with ReLU activation function. Our NID objective is formulated as a \textbf{3}-way classification task, aimed at detecting whether an input item in user sequence is shuffled, replaced randomly, or remains unchanged:
\begin{equation}\label{NTD}
\begin{aligned}
&\hat{y}_{i^{u}_{l}}^{c}=\mathrm{ReLU}(\tilde{\mathbf{h}}_{i^{u}_{l}}\mathbf{W}+\mathbf{b})\\
\mathcal{L}^{NID} = &-\frac{1}{B*L}\sum_{u=1}^{B} \sum_{l=1}^{L}  \sum_{k=0}^{2}  y_{i^{u}_{l}}^{c}  log(\hat{y}_{i^{u}_{l}}^{c}) 
\end{aligned}
\end{equation}
where $\mathbf{W}\in\mathbb{R}^{d{\times}3},\mathbf{b}\in\mathbb{R}^{1{\times}3}$ denote the weight and bias for the linear NID classifier, $y_{i^{u}_{l}}^{c}$ is the one-hot label corresponding to $3$ types of items in the corrupted user sequences.

\subsubsection{Robustness-aware Contrastive Learning} To further enhance framework robustness against data noises, we complement the NID objective, which emphasizes item-level noise adaptation, with a Robust-aware Contrastive Learning (RCL) objective to bolster representation learning ability on the user level, as is demonstrated in Figure~\ref{overview}(d). RCL makes sequence representations remain stable even when subjected to partial item shuffling or replacement, which can effectively enhance the framework's ability to handle domains with substantial stylistic differences and improve framework robustness. Specifically, for each user $u$, we employ the original user sequence $s_u$ and its corrupted user sequence $\tilde{s}_{u}$ (distorted by NID) as positive pairs, while negatives pairs are composed of other users sampled from the same batch (similar to Eq.~\ref{DAP} and Eq.~\ref{NICL_two_modalities}). The RCL objective is formulated as follows:
\begin{equation}
\begin{aligned}
\mathcal{L}^{RCL} = -\frac{1}{B}\sum_{u=1}^{B}\log& \frac{\exp(h^{\top}_u{\cdot}\tilde{h}_u)}{\exp(h^{\top}_u{\cdot}\tilde{h}_u) + \sum\limits_{1{\leq}k{\leq}B,k{\neq}u} \exp(h^{\top}_u{\cdot} \tilde{h}_k)}\\
\end{aligned}
\end{equation}
where $h_u=\mathrm{pooling}(\mathbf{h}_{i^{u}_{1}},\mathbf{h}_{i^{u}_{2}},\dots,\mathbf{h}_{i^{u}_{L}})\in\mathbb{R}^{d}$ denotes the pooled hidden representations for user sequence $s_u$ and $\tilde{h}_u=\mathrm{pooling}(\tilde{\mathbf{h}}_{i^{u}_{1}},\tilde{\mathbf{h}}_{i^{u}_{2}},\dots,\tilde{\mathbf{h}}_{i^{u}_{L}})\in\mathbb{R}^{d}$ denotes the pooled hidden representations for the corrupted user sequence $\tilde{s}_{u}$.

\subsection{Transfer Learning}\label{tranferlearning}
PMMRec is first pre-trained on the source dataset and then fine-tuned on the target dataset for transferable recommendation. The plug-and-play architectures of our framework also ensure versatility under various transfer learning settings.
\subsubsection{Pre-training on Source Data} To pre-train PMMRec on the source data, we adopt a multi-task learning strategy and sum up all the objectives as the total objective function:
\begin{equation}
\mathcal{L}=\mathcal{L}^{DAP}+\mathcal{L}^{NICL}+\mathcal{L}^{NID}+\mathcal{L}^{RCL}
\end{equation}
\subsubsection{Fine-tuning on Target Data} We directly adopt the next item prediction objective $\mathcal{L}^{DAP}$ as the only objective function when fine-tuning on the target data. This designing enables fast convergence on the target data and also ensures versatility under various transfer learning settings. 

\subsubsection{Transferring Versatility} As is shown in Figure~\ref{overview}(e), the plug-and-play architecture of each component enables PMMRec to support the following transfer learning settings:
\begin{itemize}[leftmargin = 8pt, topsep = 1pt]
    \item \textbf{Full Transfer.} This is the default setting, where the whole framework is transferred, including the text encoder, the image encoder, the multi-modal fusion module, and the user encoder. After being fine-tuned on the target data, the model can predict the next item via the probability $P(i^{u}_{L+1}|s_u)=\mathrm{softmax}(\mathbf{h}^{\top}_{i^{u}_{L}}{\cdot}\mathbf{e}^{cls}_{i^{u}_{L+1}})$.
    \item \textbf{Transfer Item Encoders.} This setting transfers the text encoder, the image encoder and the multi-modal fusion module. The probability of predicting the next item is also given by $P(i^{u}_{L+1}|s_u)=\mathrm{softmax}(\mathbf{h}^{\top}_{i^{u}_{L}}{\cdot}\mathbf{e}^{cls}_{i^{u}_{L+1}})$.
    \item \textbf{Transfer User Encoder.} This setting applies when we wish to learn general transition patterns shared across domains and platforms. In this setting, we only transfer the user encoder. The probability of predicting the next item is also given by $P(i^{u}_{L+1}|s_u)=\mathrm{softmax}(\mathbf{h}^{\top}_{i^{u}_{L}}{\cdot}\mathbf{e}^{cls}_{i^{u}_{L+1}})$.
    \item \textbf{Transfer Text Modality Only.} This setting applies when the target data have only text-modality features but lack vision-modality features. In this setting, we only transfer the text encoder and the user encoder. The text-modality feature embeddings $\mathbf{t}^{cls}_{i^{u}_{l}}$ are directly fed into the user encoder for fine-tuning, while the vision encoder and the multi-modal fusion module are discarded. The model predicts the next item via the probability $P(i^{u}_{L+1}|s_u)\!=\!\mathrm{softmax}(\mathbf{h}^{\top}_{i^{u}_{L}}{\cdot}\mathbf{t}^{cls}_{i^{u}_{L+1}})$.
    \item \textbf{Transfer Vision Modality Only.} 
    This setting applies when the target data have only vision-modality features but lack text-modality features. 
    The vision-modality feature embeddings $\mathbf{v}^{cls}_{i^{u}_{l}}$ are directly fed into the user encoder, and the model predicts the next item by $P(i^{u}_{L+1}|s_u)=\mathrm{softmax}(\mathbf{h}^{\top}_{i^{u}_{L}}{\cdot}\mathbf{v}^{cls}_{i^{u}_{L+1}})$.
\end{itemize}

For comparison on the versatility of transfer learning, we compare PMMRec with transferable recommenders and summarize the differences in Table~\ref{transferlearningsettings}. 
ID-based transferable recommenders such as PeterRec \cite{yuan2020parameter} learn user representations via ID-based interaction sequences, thus they do not apply to multi-modal settings. Transferable recommenders such as UniSRec \cite{hou2022towards} and VQRec \cite{hou2022learning} learn transferable item representations via the associated descriptions of items, thus they are only applicable to the text-only transfer learning setting. MoRec \cite{yuan2023go} demonstrates the possibility of a modality-based recommender system via a pre-trained text encoder or a pre-trained image encoder, but is not able to process recommendation data with a combination of multiple modalities. 
By comparison, our framework applies to both multi-modality and single-modality transfer learning settings and demonstrates better versatility.
\begin{table}    
    \caption{Comparison of different transfer learning settings.}
    \label{transferlearningsettings}
    \centering
    \begin{tabular}{lccccccc}
    \toprule
    \multirow{2}{*}{Methods}&\multicolumn{3}{c}{Multi-Modality}&\multicolumn{2}{c}{Single-Modality}\\ \cmidrule(lr){2-4}\cmidrule(lr){5-6}&{\thinspace}Full{\thinspace}&Item Enc.&User Enc.&Text&Vision\\
    
    \midrule
    PeterRec\cite{yuan2020parameter} &  \XSolidBrush &  \XSolidBrush &  \XSolidBrush &  \XSolidBrush &  \XSolidBrush\\
    UniSRec\cite{hou2022towards} &  \XSolidBrush &  \XSolidBrush &  \XSolidBrush &  \CheckmarkBold &  \XSolidBrush\\
    VQRec\cite{hou2022learning} &  \XSolidBrush &  \XSolidBrush &  \XSolidBrush &  \CheckmarkBold &  \XSolidBrush\\
    MoRec\cite{yuan2023go} &  \XSolidBrush &  \XSolidBrush &  \XSolidBrush &  \CheckmarkBold &  \CheckmarkBold\\
    PMMRec (ours) &  \CheckmarkBold &  \CheckmarkBold &  \CheckmarkBold &  \CheckmarkBold &  \CheckmarkBold \\
    \bottomrule
    \end{tabular}
\end{table}

\section{Experiments}
In this section, we conduct comprehensive experiments to answer the following research questions:\begin {itemize} [leftmargin = 8pt, topsep = 1pt]
\item \textbf{RQ1.} How does the proposed PMMRec, as a representation model, perform compard with the state-of-the-art baselines in sequential recommendation tasks?
\item \textbf{RQ2.} Can PMMRec outperform existing transferable sequential recommenders when used as a pre-trained model?
\item \textbf{RQ3.} Can PMMRec achieve versatility under both single-modal and multi-modal transfer learning settings?
\item \textbf{RQ4.} Will pre-training help PMMRec converge faster when fine-tuned on the target data?
\item \textbf{RQ5.} How does PMMRec perform under challenging settings such as transferring between platforms with significant style differences and the cold-start setting?
\item \textbf{RQ6.} What is the effectiveness of the proposed objectives in PMMRec?

\end{itemize}

\subsection{Experiment Settings}
\subsubsection{Datasets}
Many existing transferable recommender systems \cite{hou2022towards,hou2022learning} primarily focus on the e-commerce domain of item images with clean and pure backgrounds (e.g., items in the Amazon dataset all belong to the "fashion" category). However, there is insufficient evidence to prove whether complex multi-modal information can facilitate transfer learning in recommendation systems with complex contents (e.g., short videos featuring a wide range of topics such as food, movie and cartoon). To provide a more comprehensive evaluation, we evaluate our framework on \textbf{4} source and \textbf{10} downstream real-world datasets, namely the \textbf{HM}\footnote{https://www.kaggle.com/competitions/h-and-m-personalized-fashion-recommendations/overview} and \textbf{Amazon}\footnote{https://nijianmo.github.io/amazon/index.html} datasets for clothing and shoe purchases, and the \textbf{Bili}\footnote{https://www.bilibili.com} and \textbf{Kwai}\footnote{https://www.kuaishou.com} datasets from short-video platforms\footnote{Kwai and Bili datasets were from in previous works  \cite{zhang2023ninerec,yuan2023go}.}. For the HM and Amazon datasets, we use descriptions and covers of the products and add categorical tags to the textual features to distinguish products with the same descriptions. For the Bili\cite{zhang2023ninerec} and Kwai datasets \cite{zhang2023ninerec}, the title and the cover of each video are regarded as the multi-modal features.

We preprocess the datasets by setting the size of all images to 224 × 224, with a maximum of 50 words for all text descriptions or titles (covering more than 95\% of the descriptions). Following previous works \cite{hou2022learning,yuan2019simple}, we also filtered out users and items with fewer than five interactions in each dataset. The preprocessed dataset statistics are presented in Table~\ref{Statistic_datasets}. 

Besides, for all datasets, we consider the cold-start setting (different from above regular settings) in Section \ref{cold_start}, where we count the interactions of all items in the training set and consider those items with less than 10 occurrences as cold items. Then, we truncate the complete user sequence to get sub-sequences with a cold item at the end, all of which are used to evaluate the recommenders in the cold-start setting.

\begin{table}
\caption{Dataset statistics after preprocessing.}
\label{Statistic_datasets}
\begin{center}
  \setlength{\tabcolsep}{4pt}{
\begin{tabular}{lccccc}
\toprule
       Dataset & \#users & \#items & \#actions & avg.length & sparsity  \\
\midrule
\textbf{Source}    & 600,000   & 232,772& 6,953,503 &11.59  & 99.98\%\\
\ -Bili    & 100,000   & 44,887  &  1,537,850& 15.38  & 99.97\%\\
\ -Kwai    & 200,000   & 39,410 & 1,512,646  & 7.56   & 99.98\%\\
\ -HM    & 200,000   & 85,019  & 3,160,543  &15.80  & 99.98\%\\
\ -Amazon    & 100,000   & 63,456 &  742,464 &  7.42  & 99.98\%\\
\midrule
Bili\_Food    & 6,485     & 1,574   & 39,152  &  6.04 & 99.61\%\\
Bili\_Movie   & 16,452    & 3,493   & 114,239  & 6.94  & 99.80\%\\
Bili\_Cartoon & 30,102    & 4,702   &  211,497 & 7.03  & 99.84\%\\
\bottomrule
Kwai\_Food   & 8,549    & 2,097   &   72,741& 8.51  & 99.59\%\\
Kwai\_Movie & 8,477    & 7,024   &   60,208& 7.10  & 99.99\%\\
Kwai\_Cartoon   & 17,429    & 7,284   &  131,733 & 7.56  & 99.89\%\\
\bottomrule
HM\_Clothes            & 27,883    & 2,742   & 185,297  & 6.65  & 99.71\%\\
HM\_Shoes   & 21,666    & 3,743   &  164,621 &  7.60  & 99.81\%\\
\bottomrule
Amazon\_Clothes    & 5,009   & 5,855   &  30,383 & 6.06 & 99.89\%\\
Amazon\_Shoes           & 15,264    & 16,852  &  93,999    & 6.16 & 99.96\%\\
\bottomrule
\end{tabular}}
\end{center}
\end{table}
\begin{table*}[!htbp]
\caption{Performance comparison (\%) of different methods on source dataset. The best and the second-best performance in each row are bolded and underlined, respectively. Improvements compared with the best baseline method are indicated in the last column.}
\label{various_representation}
\begin{center}
  \setlength{\tabcolsep}{5pt}{
    \begin{tabular}{llcccccccccc}
   \toprule
   \multirow{2}{*}{Dataset} & \multirow{2}{*}{Metrics} &\multicolumn{3}{c}{IDSR} & \multicolumn{2}{c}{IDSR w. Side Feat.} & \multicolumn{3}{c}{Transferable SR}  &\multicolumn{1}{c}{Ours} &\multirow{2}{*}{Improv.} \\
   \cmidrule(lr){3-5} \cmidrule(lr){6-7} \cmidrule(lr){8-10}  \cmidrule(lr){11-11}
   &  &GRURec &NextItNet & SASRec & FDSA  & CARCA++  &UniSRec & VQRec   &MoRec++ & PMMRec \\ 
\midrule
   \multirow{6}{*}{Bili}

   &HR@10  & 3.06 & 2.66 & 4.04 &  4.46  &  \underline{5.25}   & 0.64 & 1.75
    & 4.87 & \textbf{5.49}    & +4.57\%\\
   &HR@20 & 4.37 & 4.06 & 5.76 &  6.38  &  \underline{7.98} & 1.01 & 2.69
    & 7.65 & \textbf{8.20}    & +2.68\%\\
    &HR@50 &  7.83 & 6.66 & 10.27 & 11.60  &  \underline{12.95} &  2.06 & 4.64
    & 11.54 & \textbf{13.58}    & +4.63\%\\

   &NDCG@10   & 1.57 & 1.34 &2.17 & 2.33  &  \underline{2.74} &  0.31 &  0.78
    & 2.57 & \textbf{2.90} & +5.83\%\\
   &NDCG@20  &1.85 & 1.67 & 2.61 & 2.96  &  \underline{3.34} &  0.48  &  0.92
    & 3.23 & \textbf{3.51} &  +4.84\%\\
    &NDCG@50 & 2.53 & 2.36 & 3.39&  3.86  &  \underline{4.32} &  1.62 & 1.40
    & 4.19 & \textbf{4.58}    & +5.68\%\\

   \midrule
   \multirow{6}{*}{Kwai}
   &HR@10  &4.62 & 3.69 & 5.56&  5.79  &  \underline{6.94} &  1.87 & 2.73 
   &  6.93 &\textbf{7.53} & +8.50\%\\
   &HR@20 & 6.99 &6.08 & 8.01 &  8.39  &  \underline{9.98} &  2.49 & 4.14 
   &  9.45 &\textbf{10.89} & +8.35\%\\
    &HR@50 & 11.55 &10.08 & 12.76 &  14.38  & 15.89 &  6.30 & 6.79 
   &  \underline{16.47} &\textbf{17.26} & +4.58\%\\

   &NDCG@10   &2.41 &2.33 & 2.93& 3.03  &  3.62  & 0.87 & 1.22
    & \underline{3.68}  & \textbf{4.00}  & +10.50\%\\
    &NDCG@20  & 3.00 &2.59 &3.57 & 3.96  &  4.32  &1.02 & 1.69
    & \underline{4.38}  & \textbf{4.77}  & +8.17\%\\
    &NDCG@50  & 3.91 &3.48 &4.51 & 4.89  &  \underline{5.69}  & 1.81 & 2.11
    & 5.61  & \textbf{6.03}  & +5.64\%\\
    
    \midrule
    \multirow{6}{*}{HM}

   &HR@10  & 8.39 & 8.46 & 11.60 & 11.73   &   \underline{14.65} &  3.75 & 6.25
    & 14.54 & \textbf{15.06}     & +2.80\%\\
    &HR@20 & 11.44 & 11.38 & 15.14 & 15.34   &  \underline{18.15} &  4.56 & 8.62
    & 17.98 & \textbf{19.00}     &  +4.48\%\\
    &HR@50 & 17.26 & 17.45 & 21.32 & 21.70   &  \underline{24.93} &  9.41 & 12.01
    & 24.14 & \textbf{26.34}     &  +5.35\%\\

   &NDCG@10  & 4.98 & 4.84 & 7.49 & 7.64  &  \textbf{9.63} &  1.94 & 3.33
    & 9.21 & \underline{9.54}  & -0.94\%\\
    &NDCG@20  & 5.74 & 5.54 & 8.38 & 8.52  &  \underline{10.23} &  2.31 & 3.95
    & 9.98 & \textbf{10.39}  &  +1.54\%\\
    &NDCG@50  & 6.89 & 6.87 & 9.61 & 9.95  &  \underline{11.27} &  3.16 & 4.59
    & 10.63 & \textbf{11.83}  &+4.73\%\\
    
   \midrule
   \multirow{6}{*}{Amazon}

   &HR@10 & 19.25  & 18.00 & 22.95 & 20.12  &  \textbf{23.67}  &  7.88 & 21.26 
    & 23.10  & \underline{23.57}  & -0.42\%\\
    &HR@20 & 21.15 & 19.11  & 23.17 & 21.39  &  \underline{23.83}  &  8.18 & 22.87 
    & 24.23 & \textbf{23.99}  & +0.67\%\\
    &HR@50 & 23.16 & 20.81  & 25.16 & 22.13  &  \underline{25.56}  &  9.15 & 24.49 
    & 25.87  & \textbf{26.28}  & +2.74\%\\
    
   &NDCG@10  &  17.99 & 15.59 & 20.05 &  17.82 &  \underline{20.57} & 4.69 &  15.36
    &20.61 &\textbf{20.84}  & +1.31\%\\
   &NDCG@20  & 18.29 &  15.88 & 20.32 &  18.38 & \underline{20.68} & 4.78 &  15.65
    &20.67 &\textbf{20.93}  & +1.19\%\\
    &NDCG@50  & 18.69 &  16.21 & 20.68 &  20.43 & \underline{21.02} & 4.97 &  16.07& 20.99 &\textbf{21.18}  & +0.76\%\\
    
\bottomrule
\end{tabular}}
\end{center}
\end{table*}
\subsubsection{Evaluation Metrics}
We adopt the standard \textit{leave-one-out} evaluation strategy \cite{yuan2019simple} and split the datasets into three parts: training, validation, and test data. We evaluate all models using two Top-N ranking metrics: Hit Ratio (HR@$k$) and Normalized Discounted Cumulative Gain (NDCG@$k$, or NG@$k$ for short) \cite{yuan2019simple}, where $k\in\{10,20,50\}$. Since sampled metrics might lead to unfair comparisons \cite{li2020sampled,Krichene2020sampled}, we rank the prediction results on the whole dataset.

\subsubsection{Implementation Details}
We implement our work in PyTorch \cite{paszke2019pytorch}. The hidden dimensionality $d$ is set as 768 following the configurations of the multilingual RoBERTa \cite{liu2019roberta} and the pre-trained ViT \cite{radford2021learning}. Due to limited resources, we maintain a fixed batch size of 64 per GPU, and all text and vision encoders are fine-tuned with only the top 2 Transformer blocks. We employ the AdamW optimizer \cite{loshchilov2018decoupled} and train our framework using the early stopping technique. All the pre-training experiments are carried out with 8 NVIDIA A100 GPUs and all the fine-tuning experiments are carried out with 1 NVIDIA A100 GPU.

\subsubsection{Baselines} To provide a comprehensive evaluation on the performance of our method, we compare PMMRec with the following baseline methods from three different groups\footnote{We note that previous works seldom consider the multi-modal setting involving both texts and images. For a fair comparison, we select the strongest baselines (CARCA from group 2 and MoRec from group 3), and improve them into their multi-modal versions (denoted as CARCA++ and MoRec++).}:

\noindent\textbf{Pure ID-based sequential recommenders (IDSR)} assign an unique ID to each item in the dataset. However, as item IDs are not shareable across domains and platforms, these methods are difficult to transfer to new recommendation scenarios. 
\begin{itemize}[leftmargin = 8pt, topsep = 1pt]
\item\textbf{GRURec} \cite{hidasi2015session} employs RNN as the sequence encoder and introduces ranking loss functions for model optimization.
\item\textbf{NextItNet} \cite{yuan2019simple} is a CNN-based sequential recommender, which combines masked filters with dilated convolutions to increase the receptive fields.
\item\textbf{SASRec}\cite{kang2018self} utilizes a unidirectional Transformer as the sequence encoder, which flexibly assigns attention weights to different items in the sequence.
\end{itemize}

\noindent \textbf{ID-based sequential recommenders with side features (IDSR w. side feat.)} enhance pure ID-based sequential recommenders with multi-modal side features such as texts and images. Such designing is still constrained under the ID-based paradigm, thus still falls short of transferability. 

\begin{itemize}[leftmargin = 8pt,topsep = 1pt]

\item\textbf{FDSA} \cite{zhang2019feature} captures the transition patterns between items and the contextual features of items (e.g., brands, text descriptions) via feature-based self attention blocks.
\item\textbf{CARCA++} is our improved version of CARCA \cite{rashed2022carca}. 
CARCA utilizes cross attention to capture  correlation between items and contextual features. 
The original version only supports imaged-based feature, and we improve it into a multi-modal version supporting both text and image features.
\end{itemize}

\noindent\textbf{Transferable sequential recommenders (Transferable SR)} learn transferable item representations via item texts or images. 
\begin{itemize}[leftmargin = 8pt, topsep = 1pt]
\item\textbf{UniSRec} \cite{hou2022towards} leverages the associated description text of items to learn transferable representations across different recommendation scenarios. We adopt the settings as described in the original paper.
\item\textbf{VQRec} \cite{hou2022learning} maps item texts into discrete codes and then performs the lookup operation in the code embedding table to learn transferable item representations. We adopt the settings as described in the original paper.
\item\textbf{MoRec++} is our improved version of MoRec\cite{yuan2023go}. 
MoRec adopts a pre-trained text or image encoder to extract modality embeddings and feeds the embeddings into the SASRec\cite{kang2018self} model for recommendation, which supports text or vision modality only. For fair comparison, we improve it into a multi-modal version by fusing the multi-modal representations extracted from both the text and the vision encoder.
\end{itemize}

\subsection{Overall Performance Comparisons on Source Data (\textbf{RQ1})}
We evaluate PMMRec and other baseline methods on 4 source datasets. From the experiment results in Table~\ref{various_representation}, we can reach the following conclusions:
\begin{itemize}[leftmargin = 8pt, topsep = 1pt]
    \item Compared with all the baselines, PMMRec achieves the best performance in almost every case. Notably, PMMRec even achieves performance superior than ID-based methods and on par with ID-based methods with side features. This shows PMMRec can effectively capture recommendation semantics from multi-modal contents without relying on item IDs. 
    \item Compared with other multi-modal methods (CARCA++ and MoRec++), PMMRec achieves the best performance. Different from these recommenders that lack effective representation alignment techniques, PMMRec introduces NICL to ensure that the semantics of the multi-modal representations are well aligned with the recommendation task, thereby gaining performance improvements on almost every dataset. 
    \item Compared with the best baseline CARCA++, PMMRec shows smaller performance gains on HM and Amazon datasets, but larger gains on Bili and Kwai datasets. This is probably because the former two datasets have relatively clean and pure image backgrounds, while the latter has more complex visual presentations (e.g., posters). Benefiting from the alignment and denoising strategies above, PMMRec can indeed reduce the data noises under complex semantic contexts and improve the performances, thereby showing larger performance gains on Bili and Kwai datasets.

\end{itemize}
 
\subsection{Transfer Learning Analysis (\textbf{RQ2})}
\begin{table*}
\caption{Transfer learning performance comparison (\%) on downstream datasets. w/o PT: train from scratch; w. PT: pre-train on fused 4 source datasets and then fine-tune on each downstream dataset.}
\label{compare_transfer}
\begin{center}
\begin{tabular}{llcccccccccc}
\toprule
\multirow{2}{*}{Dataset} & \multirow{2}{*}{Metric} &\multirow{2}{*}{SASRec}   &\multicolumn{2}{c}{UniSRec} 
 &\multicolumn{2}{c}{VQRec}  &\multicolumn{2}{c}{MoRec++} &\multicolumn{2}{c}{PMMRec}  & \multirow{2}{*}{Improv. (\%)}\\
\cmidrule(r){4-5}\cmidrule(r){6-7}  \cmidrule(r){8-9}  \cmidrule(r){10-11}
    & &  &w/o PT &w. PT  &w/o PT &w. PT   &w/o PT &w. PT &w/o PT &w. PT\\
\midrule
%\midrule
\multirow{2}{*}{Bili\_Food}
&HR@10 & 16.55  &2.21 & 7.40 & 14.96 &17.61 & 18.67&  19.09  & \underline{20.05} & \textbf{22.67}      & +13.06\%\\ 
&NDCG@10 &9.81  & 1.16 & 3.43 &7.06 &8.62 & 9.15 & \underline{10.86} & 10.80 & \textbf{12.75}  & +14.82\%\\ 
\midrule
\multirow{2}{*}{Bili\_Movie}
&HR@10 & 11.60 & 5.38& 6.78&10.23 &11.09 & 12.04 &12.69 & \underline{13.50} & \textbf{15.02} & +11.26\%\\ 
&NDCG@10  & 6.54 &2.50 & 3.25 &5.08 &5.49 & 6.17 & 7.01 & \underline{7.45} & \textbf{8.35}  & +12.08\%\\ 
\midrule
\multirow{2}{*}{Bili\_Cartoon}
&HR@10 & 11.59  & 3.66 &5.37 &10.14 &10.97 & 12.64 &13.76 &\underline{14.49} & \textbf{15.82}     & +9.18\%\\ 
&NDCG@10  & 6.66 &1.86 & 2.82 &4.63 &5.50 & 6.99 & 7.87 & \underline{8.10} & \textbf{8.96}  & +10.62\%\\ 
\midrule

\multirow{2}{*}{Kwai\_Food}
&HR@10 & 33.17 & 23.84 & 9.21$\downarrow$ &25.84 &26.21 & 31.76 & 33.72 &  \underline{37.03} & \textbf{38.51}  & +4.00\%\\ 
&NDCG@10  & 18.32 &12.33 & 4.30$\downarrow$  &11.41 &12.51 & 16.98 & 18.02  & 19.51 & \textbf{20.79}  & +6.56\%\\ 
\midrule

\multirow{2}{*}{Kwai\_Movie}
&HR@10 &  6.08 & 0.92 & 2.56& 4.51 & 4.22$\downarrow$ & 5.07& 6.86 & \underline{7.43} &  \textbf{8.84}     & +18.98\%\\ 
&NDCG@10  & 3.32 &0.47 & 1.28 &2.08 &2.03$\downarrow$ & 2.92 & 3.43 &  \underline{3.86} & \textbf{4.88}  & +26.43\%\\ 
\midrule
\multirow{2}{*}{Kwai\_Cartoon}
&HR@10 &  12.87 & 8.74 & 4.62$\downarrow$ &10.52 & 9.54$\downarrow$ & 10.39 & 11.92& \underline{15.39} & \textbf{16.42}  & +6.69\%\\ 
&NDCG@10  &7.34  &4.10 & 2.19$\downarrow$ &5.03 & 4.47$\downarrow$ &5.17 & 6.57 & \underline{8.26} & \textbf{8.98}  & +8.71\%\\ 
\midrule
\multirow{2}{*}{HM\_Clothes}
&HR@10 & 9.94 &3.57 & 6.78 &8.92 &9.52 &  10.51  & \underline{11.75}  &  10.13 & \textbf{14.70 }      & +20.07\%\\ 
&NDCG@10 &5.13 &1.91 &  3.30 &3.88  &4.72  & 5.56 & \underline{6.24} & 4.98 &  \textbf{7.98}  & +21.81\%\\ 
\midrule
\multirow{2}{*}{HM\_Shoes}
&HR@10 & 13.99 & 9.22 & 7.28$\downarrow$ &11.70 & 12.03  & 12.36& \underline{14.94}& 14.30 &  \textbf{18.97}      & +21.24\%\\ 
&NDCG@10 & 8.26 &4.79 & 3.67$\downarrow$  &5.46  &6.21 & 6.85 & \underline{8.52} & 7.86 & \textbf{11.07}  &  +23.03\%\\ 
\midrule
%\midrule
\multirow{2}{*}{Amazon\_Clothes}
&HR@10 & 40.71 & 34.94 &  36.44 &  40.32 & 40.77&  37.67 &   40.09    & \underline{40.42}  & \textbf{43.78}  & +8.31\%\\ 
&NDCG@10  & 39.30  & 19.48 &  22.28 & 22.17 & 22.68 & 36.17  & 38.41  &  \underline{38.97} & \textbf{41.09}  & +5.44\%\\ 
\midrule
\multirow{2}{*}{Amazon\_Shoes}
&HR@10 &  11.80    & 6.47&  7.07 & 12.79 & 12.74 &   12.97  &  \underline{13.46}&  11.85  &  \textbf{15.97}  & +34.77\%\\ 
&NDCG@10  &  9.39 & 5.19  & 4.62  & 8.41 & 8.80  & 8.56 &  \underline{9.43} & 9.14 &   \textbf{13.10}   & +43.32\%\\ 
\bottomrule
\end{tabular}
\end{center}
\end{table*}
Following \cite{hou2022towards,hou2022learning}, we pre-train PMMRec and other transferable recommenders on fused 4 source datasets, then fine-tune on each downstream cross-platform dataset respectively to evaluate their transfer learning performances. For comparison, we also directly train each recommender on each downstream dataset from scratch without pre-training, and report their performances. Based on the experiment results in Table \ref{compare_transfer}, we can reach the following conclusions:
\begin{itemize}[leftmargin = 8pt, topsep = 1pt]
    \item Both PMMRec and MoRec++ outperform UniSRec and VQRec on all datasets by a large margin. This is probably because PMMRec and MoRec++ can effectively capture the complex transition patterns between multiple modalities, thus showing better performance than transferable methods relying on item texts only (e.g., UniSRec and VQRec)\footnote{We observe that UnisRec and VQRec do not perform well under our experimental setting, which might be attributed to the following reasons: non-end-to-end training manner~\cite{yuan2023go}, lack of multimodal information processing modules, and the high semantic complexity of our datasets, particularly the Kwai and Bili datasets.}.
    \item We notice that PMMRec is superior to MoRec++, which verifies that by introducing NICL to achieve representation alignment, our PMMRec can learn more expressive multi-modal representations that are suitable for the recommendation task, hence outperforming other baseline methods that are not equipped with effective alignment techniques.
    \item Under complex scenarios (e.g., Bili and Kwai), PMMRec manages to stay robust against data noises and content differences owing to the introduction of the self-supervised denoising objectives, while other methods fail to handle the content differences and exhibit significant declines.
    \item  Although transfer learning can leverage upstream knowledge to improve recommendation performance on the target dataset, we notice that UniSRec performs significantly worse than the ID-based SASRec, and VQRec achieves comparable performance with SASRec only when pre-trained. This finding confirms the superiority of the ID-based paradigm over the past few decades. However, with more advanced transferable methods such as MoRec++ and our PMMRec, knowledge from other platforms can be more effectively utilized, providing researchers with the opportunity to further refreshing the upper bound of recommendation performance.
\end{itemize}
\subsection{Versatility of Transfer Learning (\textbf{RQ3})}
As we have mentioned above, PMMRec can achieve versatility under various transfer learning settings. To verify the versatility of PMMRec, we examine 5 typical transfer learning settings, including two single-modality transferring settings: (1) transfer text modality only (PMMRec-T w. PT); (2) transfer vision modality only (PMMRec-V w. PT), and three multi-modality transferring settings: (3) transfer item encoders (PMMRec-I w. PT); (4) transfer user encoder (PMMRec-U w. PT); (5) full transfer (PMMRec w. PT). For comparison, we also directly train PMMRec on each downstream dataset without pre-training (PMMRec-T w/o PT: train PMMRec from scratch with only text data; PMMRec-V w/o PT: train PMMRec from scratch with only vision data; PMMRec w/o PT: train PMMRec from scratch with multi-modal data). The performances of PMMRec under different settings are presented in Table \ref{various_transfer}. From Table \ref{various_transfer}, we find that although PMMRec is pre-trained with multi-modal data, the cross-modal alignment and fine-grained fusion strategies enable PMMRec to handle both multi-modal and single-modal settings. Compared to the multi-modal setting (PMMRec w. PT), PMMRec under the single-modal settings (PMMRec-T w. PT, PMMRec-V w. PT) still manages to maintain competitive performances. 

Furthermore, we find that the full transfer of both the item encoders and the user encoder (PMMRec w. PT) achieves the best performance. Also, for single-architecture transfer, transferring the item encoders (PMMRec-I w. PT) outperforms transferring the user encoder (PMMRec-U w. PT) in most cases. Transferring the user encoder only leads to performance improvements on very few datasets, but transferring the item encoders can achieve competitive performances compared to full transfer. These findings suggest that, by aligning item representations with recommendation semantics, the NICL objective manages to incorporate the transition patterns of user behaviors into the item encoders, thus, to some extent, empowering the item encoders with the recommendation ability. 
\subsection{Convergence Analysis (\textbf{RQ4})}
\begin{figure*}[htbp]
    \begin{center}
        \includegraphics[width=1.0\linewidth]{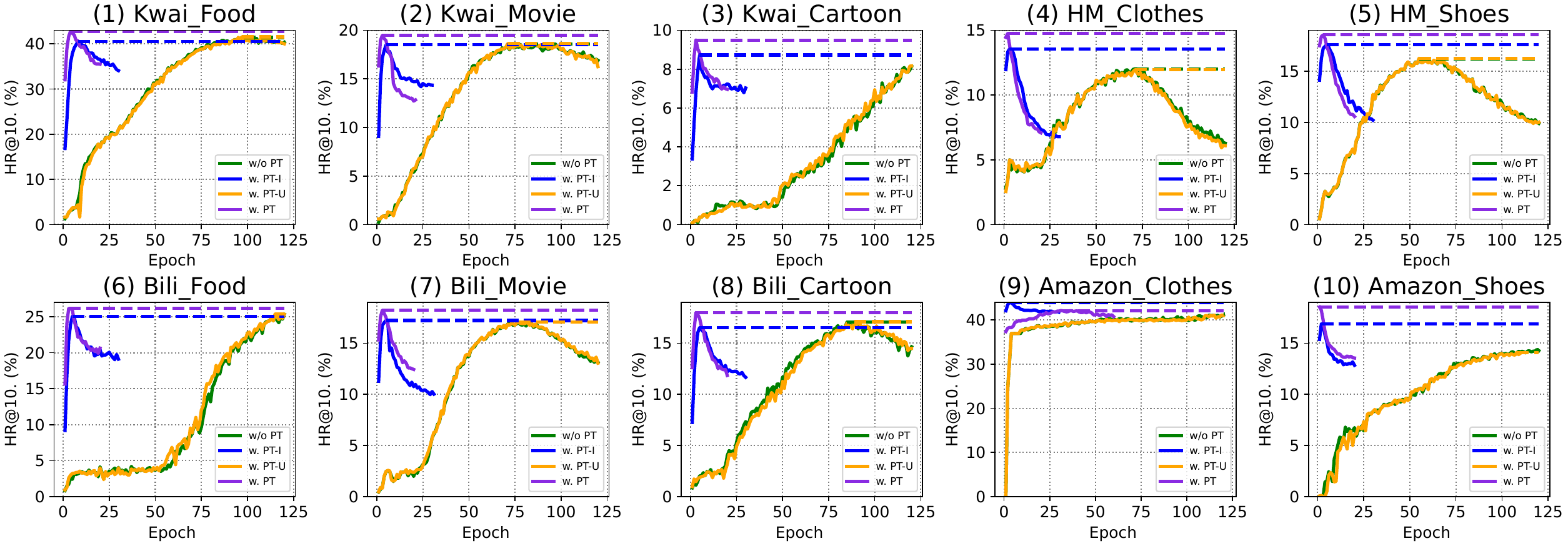}
    \end{center}
        \caption{Convergence curves on downstream datasets under different transfer learning settings. w/o PT: train from scratch; w. PT-I: transfer item encoders; w. PT-U: transfer user encoder; w. PT: full transfer.}
    \label{transfering}\vspace{-0.5cm}
\end{figure*}

\begin{table}
\caption{Performances (\%) under versatile transfer learning settings.}%\vspace{-0.24cm}
\label{various_transfer}
\begin{center}
\fontsize{6.5}{7}\selectfont
\setlength{\tabcolsep}{1.5pt}{
\begin{tabular}{llcccccccc}

\toprule
\multirow{2}{*}{Dataset} & \multirow{2}{*}{Metric}   &\multicolumn{2}{c}{PMMRec-T}  &\multicolumn{2}{c}{PMMRec-V}  &\multicolumn{4}{c}{PMMRec} \\
\cmidrule(r){3-4} \cmidrule(r){5-6} \cmidrule(r){7-10} 
     &  &w/o PT &w. PT  &w/o PT &w. PT   &w/o PT &w. PT-I &w. PT-U &w. PT\\  
\midrule
%\midrule
\multirow{2}{*}{Bili\_Food }
&HR@10  & 19.12  & 19.26 &17.45 & 17.81	&20.05  & 20.10 & \underline{20.16} & \textbf{22.67} \\
&NG@10 & 9.34  & 10.90& 8.78 & 10.12 &10.80  & \underline{11.50} &  11.02   &\textbf{12.75}\\
\midrule
\multirow{2}{*}{Bili\_Movie}
&HR@10  &12.45 & 10.66$\downarrow$ & 10.98 &11.99 & \underline{13.50} & 13.23$\downarrow$  & 13.83 & \textbf{15.02}\\
&NG@10 & 6.96 & 5.77$\downarrow$ &  6.03 & 6.69 & 7.45  & \underline{7.68} &  7.65  & \textbf{8.35}\\
\midrule
\multirow{2}{*}{Bili\_Cartoon}
&HR@10  &13.43  & 13.57 & 11.65 & 11.14$\downarrow$& 14.49 & 14.63 & \underline{14.72} & \textbf{15.82}\\
&NG@10 &7.78  & 7.97 & 6.54 & 6.24$\downarrow$ & 8.10  & \underline{8.27}  & 8.25  & \textbf{8.96}\\
\midrule
%\midrule
\multirow{2}{*}{Kwai\_Food}
&HR@10 &  32.20  &30.63  &33.21 & 33.00$\downarrow$  & 37.03 &\underline{37.24} &  36.89$\downarrow$   & \textbf{38.51}\\
&NG@10 &17.13  &15.35  & 17.69 & 17.86 & \underline{19.51}  & 19.39$\downarrow$ & 19.27$\downarrow$ & \textbf{20.79}\\
\midrule
\multirow{2}{*}{Kwai\_Movie}
&HR@10 & 6.85   &7.23 &6.29 & 6.22$\downarrow$  & 7.43 & \underline{8.06} & 7.84 & \textbf{8.84}\\
&NG@10 & 3.47 &3.89 & 3.21 & 3.43 & 3.86 &  \underline{4.46} & 4.05  & \textbf{4.88}\\
\midrule
\multirow{2}{*}{Kwai\_Cartoon}
&HR@10& 14.27 &14.18$\downarrow$ &13.67 &  12.31$\downarrow$ & 15.39 & \underline{15.44} & 15.02$\downarrow$  & \textbf{16.42}\\
&NG@10 & 7.86 & 7.44$\downarrow$ & 7.12& 6.96$\downarrow$ & 8.26  & \underline{8.36}   & 8.00$\downarrow$  & \textbf{8.98}\\
\midrule
%\midrule
\multirow{2}{*}{HM\_Clothes}
&HR@10 & 8.57  & 11.86  &9.12  &12.02 & 10.13 & \underline{13.39} & 11.41 & \textbf{14.70}\\
&NG@10&  4.03  & 6.30 & 4.24 & 6.33 & 4.98 &  \underline{7.19} & 5.92  & \textbf{7.98}\\
\midrule
\multirow{2}{*}{HM\_Shoes}
&HR@10 & 12.98  & 16.30 & 13.17 & 15.92 & 14.30 & \underline{17.54} &  16.27  & \textbf{18.97}\\
&NG@10 & 7.26  & 9.55 & 7.48 &9.04  & 7.86 & \underline{10.25} & 9.45 & \textbf{11.07}\\
\midrule
%\midrule
\multirow{2}{*}{Amazon\_Clothes}
&HR@10 &  40.23 & 40.73 &39.98  & 41.84  & 40.42 &\underline{40.47}	 &  39.11$\downarrow$ & \textbf{43.78}\\
&NG@10 & 38.65  &  38.87 & 38.48 &39.94 &\underline{38.97}  &38.74 & 37.26$\downarrow$  & \textbf{41.09}\\
\midrule
\multirow{2}{*}{Amazon\_Shoes}
&HR@10  &11.34   & 12.83 & 11.46 &  14.65 & 11.85 & \underline{12.07} &  12.64 & \textbf{15.97}\\
&NG@10  & 8.99 & 10.97 &  8.94 & 12.20 &  9.14  &10.07 & \underline{10.81}  & \textbf{13.10}\\
%\midrule
\bottomrule
\end{tabular}}
\end{center}
\end{table}

\begin{table}
\caption{Single source transfer learning performances (\%).}
\label{single_domain}
\fontsize{6.5}{7}\selectfont
\begin{center}
\setlength{\tabcolsep}{3pt}{
\begin{tabular}{llcccccc}
\toprule
\multirow{2}{*}{Dataset} & \multirow{2}{*}{Metric}  & \multicolumn{2}{c}{No Source}  & \multicolumn{4}{c}{Source}   \\
\cmidrule(r){3-4}   \cmidrule(r){5-8}  
     &  &ID & w/o PT  &Bili  &Kwai  &HM &Amazon\\  
\midrule
%\midrule
\multirow{2}{*}{Bili\_Food}
& HR@10  &  16.55  &  20.05 &  \textbf{22.07} &   20.63 &  19.98 $\downarrow$ &  19.78 $\downarrow$\\
& NG@10  &  9.81   &  10.80 & \textbf{12.37}  & 11.61   &11.23  &  11.18 \\
\midrule
\multirow{2}{*}{Bili\_Movie}
& HR@10  &   11.60 & 13.50 & \textbf{14.59} &   13.94 & 13.23  $\downarrow$ &  13.62   \\
& NG@10  &  6.54  &  7.45 & \textbf{8.01} &  7.69 & 7.42 $\downarrow$ &  7.51  \\
\midrule
\multirow{2}{*}{Bili\_Cartoon}
& HR@10  &  11.59  &  14.49 &  \textbf{15.66} & 14.68 & 14.33 $\downarrow$ &  14.31  $\downarrow$   \\
& NG@10  &  6.66   &  8.10 & \textbf{8.83}  &  8.30   & 8.08 $\downarrow$   & 7.92 $\downarrow$ \\
\midrule
\multirow{2}{*}{Kwai\_Food}
& HR@10 &  33.17  &  37.03  & 37.36 &   \textbf{39.06} & 36.72 $\downarrow$     & 36.69 $\downarrow$\\
& NG@10&  18.32  &  19.51  & 20.06 & \textbf{21.00}   & 19.54  &  19.97  \\
\midrule
\multirow{2}{*}{Kwai\_Movie}
& HR@10 &   6.08  &  7.43 & 6.90 $\downarrow$&   \textbf{9.19}  & 6.16 $\downarrow$  &   6.39  $\downarrow$ \\
& NG@10 &  3.32  & 3.86 & 3.53 $\downarrow$ &  \textbf{4.74}   & 3.39 $\downarrow$  &  3.36  $\downarrow$ \\
\midrule
\multirow{2}{*}{Kwai\_Cartoon}
& HR@10&  12.87  &  15.39  & 13.72 $\downarrow$ &   \textbf{15.94}      & 13.51 $\downarrow$   &   14.06 $\downarrow$ \\
& NG@10&  7.34  &  8.26 & 7.54 $\downarrow$ &  \textbf{8.63}  & 7.39 $\downarrow$  &  7.57 $\downarrow$ \\
\midrule
\multirow{2}{*}{HM\_Clothes}
& HR@10 &  9.94  &  10.13 &  11.60 &  12.41  & \textbf{14.48}     &  11.36 \\
& NG@10 & 5.13    &  4.98 & 6.02 &  6.54 & \textbf{7.77}  &  5.81  \\
\midrule
\multirow{2}{*}{HM\_Shoes}
& HR@10 &  13.99  &  14.30 & 16.13 &  16.85 & \textbf{18.74}  &   15.40  \\
& NG@10 &  8.26  &  7.86 & 8.94   &  9.52  & \textbf{10.97}  &  8.43  \\
\midrule
\multirow{2}{*}{Amazon\_Clothes}
& HR@10 &  40.71  &  40.42 &  40.59 &  40.35  & 41.01  &  \textbf{43.56}  \\
& NG@10 & 39.30   &  38.97 & 39.09&  39.12    & 39.83  &   \textbf{41.01}  \\
\midrule
\multirow{2}{*}{Amazon\_Shoes}
& HR@10  &  11.80   &  11.85  & 13.73    &   13.69  & 13.48  & \textbf{ 14.45}  \\
& NG@10  &  9.39  &  9.14 & 11.68 &  11.72 & 11.58   & \textbf{12.20} \\
\bottomrule
\end{tabular}}
\end{center}
\end{table}
\label{Visualizations}

To investigate whether pre-training helps PMMRec converge faster when fine-tuned on downstream datasets, we plot the convergence curves of PMMRec under different transfer learning settings in Figure \ref {transfering}. We can see that pre-training not only substantially boosts performance but remarkably accelerates convergence speed, achieving the best performance within the first few epochs and yielding speed acceleration ranging from tens to hundreds of times. 

Interestingly, we find out that (1) transferring the item encoders (w. PT-I) approximates the performance of full transfer (w. PT) at least in terms of convergence speed; (2) transferring the user encoder alone (w. PT-U) shows marginal improvement but transferring both the user encoder and the item encoders (full transfer, w. PT) achieves significant performance improvement. The fact that transferring item encoders achieves performances comparable with full transfer probably suggests a pioneering exploration for transferable RecSys: empowering the item encoders with strong recommendation ability and making the user encoder lightweight, which is akin to the pre-trained foundation models in NLP and CV (e.g., the simple MLP classification head in BERT\cite{devlin2018bert}, the asymmetric designing of the encoder-decoder architecture in MAE \cite{He_2022_CVPR}) .

\subsection{Performance under Challenging Settings (\textbf{RQ5})}
\subsubsection{Single-source and Cross-platform Transfer Learning}
In contrast to the previous setting, we evaluate PMMRec under a different transfer learning setting, where we pre-train PMMRec on a \textbf{single} source dataset, and fine-tune PMMRec on 10 downstream datasets separately. Note that this setting is very challenging, since there are significant gaps between two completely different platforms. The results in Table \ref{single_domain} show that PMMRec achieves the best performances when transferring to homogeneous datasets (diagonal part marked in bold in Table \ref{single_domain}), and the second-best performance when transferring to heterogeneous datasets. 

Surprisingly, we find out that PMMRec pre-trained on Bili and Kwai (complex scenarios) manages to maintain good performances (lower-left part in Table \ref{single_domain}, such as Bili $\rightarrow$ HM\_Shoes) after transferring to HM and Amazon (simple scenarios) datasets. This indicates that PMMRec can overcome content differences and learn transition patterns general enough to transfer across domains and platforms. However, when transferring from simple scenarios to complex scenarios (particularly on Kwai), the performances cannot be sustained (upper-right part in Table \ref{single_domain}, such as HM $\rightarrow$ Kwai\_Movie). This phenomenon reveals that it is rather challenging to make accurate recommendations under complex scenarios.

\subsubsection{Performance under the Cold-start Setting}
\label{cold_start}
We evaluate SASRec, PMMRec-T and PMMRec-V, and PMMRec in the cold-start setting of the \textbf{4} source datasets. As is shown in Table \ref{tb: cold}, PMMRec, PMMRec-T and PMMRec-V all significantly outperform the ID-based SASRec when predicting the cold-start items. The reason may be that SASRec learns the representations only from the behavior patterns while PMMRec can additionally model the raw modality features, and this advantage of PMMRec is magnified in the cold-item setting. When compared to instance-level item representations provided by item IDs, the fine-grained tokens of texts or patches of images in PMMRec can effectively expand the range of representation space and improve expressiveness. It is also worth noting that PMMRec-T outperforms PMMRec-V in cold-start situations, which is probably due to the difference of information density between texts and images \cite{He_2022_CVPR}. 

\begin{table}
   \caption{Performance comparison (\%) under the cold-start setting.}
   \begin{center}
\fontsize{6.5}{7}\selectfont
\setlength{\tabcolsep}{5pt}{
   \begin{tabular}{llcccc}
   \toprule
   \multirow{1}{*}{Dataset} & \multirow{1}{*}{Metrics} & SASRec &PMMRec-T &PMMRec-V &PMMRec \\
    \midrule
    %\midrule

   \multirow{2}{*}{Bili}
   &HR@10
   & 0.0883  &\textbf{1.1476}  &0.6886  &1.0240\\
   &NDCG@10 
   & 0.0333  &\textbf{0.5720} &0.3486  & 0.5632\\
   \midrule
   
   \multirow{2}{*}{Kwai}
   &HR@10
   & 0.0311 &2.9490  &2.9191  &\textbf{3.5106}\\
   &NDCG@10 
   & 0.0333  &1.5139 &1.4622  &\textbf{1.7448}\\

    \midrule
   \multirow{2}{*}{HM}
   &HR@10
   & 0.0576 & \textbf{2.1767}  &1.3893 &2.0387\\
   &NDCG@10 
   & 0.0301 &\textbf{1.2931} &0.7478 &1.1659 \\
   \midrule
\multirow{2}{*}{Amazon}
   &HR@10
   & 0.1276 & 3.6437  &3.3248 &\textbf{4.1646}\\
   &NDCG@10 
   & 0.1023 &3.5998 &3.2429 &\textbf{4.0034} \\
   \bottomrule
    \end{tabular}}
    \end{center}
   \label{tb: cold}
\end{table}

\subsection{Ablation Study (\textbf{RQ6})}
PMMRec adopts an NICL objective to achieve cross-modal representation alignment and two self-supervised denoising objectives (NID and RCL) to enhance framework robustness. To verify the effectiveness of each objective, we conduct an ablation study and report the experiment results in Table~\ref{various_components}. 

For the cross-modal contrastive learning objective, we design three variants: (1) removing NICL; (2) replacing NICL with VCL; (3) replacing NICL with NCL. Comparing (1) with PMMRec, we can see that disabling NICL leads to suboptimal performance, indicating the necessity of achieving representation alignment between muli-modalities and representation alignment in the recommendation semantic space. Comparing (2) with PMMRec, we can see that excluding intra-modality negative samples results in performance decrease, which suggests that both inter- and intra-modality samples are necessary to achieve alignment between multi-modal representations. Comparing (3) with PMMRec, we can see that removing the next-item positive samples leads to performance degradation, since in such case the representations of the item encoders cannot be well aligned with recommendation semantics. 

For the self-supervised denoising objectives, we design two variants: (4) removing NID; (5) removing RCL. Comparing (4)(5) with PMMRec, we find that both the NID and the RCL play a crucial role in enhancing framework robustness against noises, and removing any objective undermines performance.
\begin{table}

\caption{Ablation study (\%) on the objective functions in PMMRec.}\vspace{-0.21cm}
\label{various_components}
\begin{center}
\fontsize{6.5}{7}\selectfont
\setlength{\tabcolsep}{2pt}{
   \begin{tabular}{llcccccc}
   \toprule
   Dataset&Metrics&w/o NICL&only VCL&only NCL& w/o NID & w/o RCL &PMMRec  \\
   \midrule
   \multirow{2}{*}{Bili\_Movie}
   &HR@10
    & 14.24 & 14.86 & 14.55  & 14.76  & 14.81 & \textbf{15.02}  \\
   &NG@10 
    & 7.56 & 7.88 & 7.96 & 8.01  & 8.13 & \textbf{8.35}  \\
   \midrule
   \multirow{2}{*}{Kwai\_Movie}
   &HR@10
    & 7.74 & 7.68 & 8.15 &  8.44 &  \textbf{8.93} &  8.84 \\
   &NG@10 
     & 4.01 & 4.14 & 4.43 &  4.65 & 4.80 &  \textbf{4.88}  \\
    \midrule
    \multirow{2}{*}{HM\_Shoes}
   &HR@10
     &13.01 & 12.67 & 13.95 & 14.21 &  14.52 &  \textbf{14.70} \\
   &NG@10 
    & 6.25 & 6.19  &  7.01 &  7.74 &  7.84&   \textbf{7.98}\\
   \midrule
   \multirow{2}{*}{Amazon\_Shoes}
   &HR@10
     & 39.13  & 40.80 & 42.24  & 42.25 &  43.83 &   \textbf{43.98} \\
   &NG@10 
    & 38.24 & 39.82 & 40.30 & 40.65 & 40.55  &   \textbf{41.09} \\
\bottomrule
\end{tabular}}
\end{center}
\end{table}
\section{Conclusion}
In this paper, we propose PMMRec, a pure multi-modality based recommender system capable of achieving performances superior or on par with ID-based recommenders under versatile transfer learning settings. Instead of item IDs, PMMRec leverages the multi-modal contents of items to learn transferable recommendation representations via a plug-and-play framework architecture. To align the cross-modal item representations, we propose an NICL objective, which is equipped with both inter- and intra-modality negative samples and explicitly incorporates the transition patterns of user behaviors into the item encoders. To enhance framework robustness against data noises, we propose an NID objective to denoise the items and a RCL objective to ensure the stability of sequence representations. Extensive experiments on real-world datasets verify the effectiveness of our approach. Looking forward, we hope to improve the generality of PMMRec by adapting PMMRec to more recommendation tasks such as rating prediction and multi-behavior recommendation, thereby paving the way for a more generic RecSys foundation model.

\section*{Acknowledgments}
This work is partially supported by the NSF of China(under Grants U22A2028), CAS Project for Young Scientists in Basic Research(YSBR-029) and Youth Innovation Promotion Association CAS.
\bibliographystyle{IEEEtran}
\bibliography{Refrences}

\end{document}